\documentclass[12pt]{article}
\usepackage{graphicx,amsmath,amssymb}
\newcommand{\func}[1]{\text{#1}\,}
\begin{document}

\title{Hydrogen induced nonmonotonic relaxation in binary mixtures
similar to Pd--Er alloys as a transition process in nonequilibrium heterogeneous 
systems with spinodal decomposition} 
\author{A.\,A. Katsnelson\footnote{albert@solst.phys.msu.su}, 
I.\,A. Lubashevskiy, A.\,Yu. Lavrenov\\
\textit{Department of Solid State Physics,} \\ 
\textit{Faculty of Physics, M.V.Lomonosov Moscow State University,} \\
\textit{119899, Moscow, Russia}} 
\date{\today}
\maketitle

\begin{abstract}

We have proposed a qualitative model for the structure of binary systems 
similar to Pd--Er alloys, which explains their nonmonotonic relaxation
after the hydrogen saturation. It is based on the assumption that such a solid 
solution involves two kind heterogeneities. The former are caused by 
spinodal decomposition of the initially homogeneous state of the solid solution
into the phases enriched and depleted of Er atoms. The latter are 
crystalline defects that trap an additional amount of Er atoms, which
leads also to their local accumulation, changing the defect properties. Hydrogen atoms penetrating
into the solid disturb the equilibrium of both the phase separation
and the defect saturation with Er atoms, causing redistribution
of Er atoms. The diffusion fluxes give rise to the motion of the
interface between the two phases that is responsible for time variations,
for example, in the relative volume of the enriched phase observed
experimentally. We have found the conditions when the interface motion can
change the direction during the system relaxation to a new equilibrium 
state. The latter effect is, from our point of view, the essence of
the hydrogen induced nonmonotonic relaxation observed in such systems.
The numerical simulation confirms the basic assumptions.

\end{abstract}

\section{Introduction}

Metal-hydrogen systems are singled out with the fact that they can
remain in nonequilibrium states for a long time because of the 
ability of hydrogen atoms to migrate relatively fast through metallic matrices. Therefore, 
a metallic matrix initially saturated with hydrogen inevitably loses its main amount
during a certain time period which may be
sufficiently long \cite{1,2,3}. Hydrogen atoms penetrating into the
metallic matrix cause the reformation of the defects structure
and as a result, the system structure becomes substantially
heterogeneous, moreover, it can be of multiscale geometry \cite{4,5}. 
When the heterogeneity scales expand over several levels these systems 
may be treated as fractals (see, e.g., \cite{6a,6b,6c}) and their 
evolution goes in a sophisticated manner. In particular, under certain
conditions \cite{7} the dynamics of such a systems is described by
strange attractor, which reflects in nonmonotonic or even chaotic time
variations in its structure and occurring phase transitions \cite{8}.     

The characteristics of structure transformations during the hydrogen
saturation and the following relaxation have been studied for metallic
alloys based on Pd \cite{8,9,10,12,13}. For example, for the Pd--Er 
alloys undergone deformation the hydrogen saturation gives rise to the
change in the sign of the elastic tension at the initial stage of
relaxation \cite{13}. This effect is due to variations in 
the image forces 
that are induced by the transformation of the defect-metal (D--M)
complexes into the hydrogen-defect-metal (H--D--M) ones because of  
the high binding energy 
between hydrogen atoms and defects in Pd matrix \cite{14}. Having
changed the sign, the tensions grow during two days, then, drop and decrease by 25\% in 
eight days. In parallel with it oscillatory
variations in the relative volume of the phases enriched and
depleted of Er atoms occur practically right after the hydrogen
saturation. The difference in the concentration of Er atoms between the
two phases evolves in the same manner. These oscillations are chaotic
rather than regular \cite{8,10}. 

The given phenomena were observed in such Pd--Er alloys where the atomic
concentration of Er atoms was about 10\% \cite{10}. So their
structure has to be sufficiently simple, namely, should comprise only
two phases formed of the Pd matrix depleted of or enriched with Er
atoms. So the relaxation after the hydrogen saturation seems to go only
through time variations in the relative volume of these phases without 
complex structural transformations in the Pd--Er system  that should appear 
for high values of the Er concentrations \cite{19}. 

The stochastic behavior of the relaxation process in the Pd--Er--H system has
been demonstrated in a frameworks similar to the Lorenz model
proposed for analysis of turbulent phenomena in atmosphere \cite{15}
and, then, applied to the description of plastic deformation
\cite{16}. It should be pointed out that time variations in the structure 
of the given alloys after the hydrogen saturation enable us to consider 
diffusion flux in them also turbulent. It can be justified at least for 
some stages of the relaxation if we take into account not only the 
stochastic evolution of the alloy structure but also the difference in 
the rate of nonmonotonic structure transformations for different
coherent diffraction regions,
as it was noted for the first time in \cite{12}. The analysis
presented in \cite{8} has been carried out in the frameworks of
macroscopic approach, so, in order to complete it, we should develop 
a microscopic mechanism by which these phenomena can be implemented. The latter 
problem is actually the purpose of the present paper.

Papers \cite{8,12} demonstrate that the macroscopic model to be
developed for the hydrogen induced relaxation in alloys similar to 
Pd--Er--H has to take into account the following. 

\begin{enumerate}
\item 
In such systems the hydrogen saturation initiates several interacting
processes, in particular, the transformation of the D--M complexes into 
D--M--H 
ones and the diffusion of Er atoms between the enriched and depleted
phases of the Pd-matrix. These two processes cause nonmonotonic time
variations not only in the relative volume of these phases but also in the
Er concentration inside them.    

\item The correlated variations in the Er concentration and the
phase relative volume suggest that the two processes stem from the same
phenomenon. Besides, their nonmonotonicity points out that diffusion of 
Er atoms does not obey conventional Fick's low, at least, everywhere in
the alloy.

\item The D--Er--H complexes induced by the hydrogen saturation have to be
enriched not only with hydrogen atoms but also with Er atoms because
of high affinity of H and Er atoms. So we meet four type regions (fig. 2): 
two regular phases of the Pd-matrix
and regions with two type defects D--Er--H (inside each of the
phases). Besides, substantial difference in the specific
volumes of the regular phases and complexes can lead to the appearance
of transitional regions between the matrix and complexes.   

\item The complexes are the source of additional nonuniformity in the
distribution of Er atoms, causing an additional diffusion flux in the
system.      

\item The variations in the phase relative volume mean displacement 
of the phase boundaries among with diffusion of Er atoms. So dealing
with the nonmonotonic relaxation on systems similar to Pd--Er alloys we
first of all should explain why a nonmonotonic motion of the phase
boundaries occur.

\end{enumerate}

It should be pointed out that the system Pd--Er--H is not unique keeping 
in mined the aforesaid properties. Even more pronounced chaotic
behavior of the structural evolution has been observed in the Pa--Ta--H
system \cite{12}, which suggests the multiscale organization of its
structure and the substantially nonequilibrium state of the defects
and the adjacent regions of the Pd-matrix. 
    
In spite of the available number of theoretical models cited above, 
physics of the hydrogen induced relaxation in such  metallic
systems is far from being understood well. So   
at the first step of developing a microscopic model able to explain
the aforesaid phenomena, it is reasonable to confine ourselves to the
analysis of the interaction between the D--M--H complexes and the phase
boundaries through diffusion of Er atoms. In this study we should also 
take into account that the
distribution of Er atoms around the D--M--H defect can be nonmonotonic
because under certain conditions the deformation field in their
neighborhood exhibits a nonmonotonic behavior. Nevertheless in the
present paper we consider only defects with a sufficiently simple
structure and propose a model which qualitatively explains the
nonmonotonic motion of the phase boundaries due to the hydrogen
saturation. We do not claim on a sophisticated theory of this
relaxation process but only show the feasibility of its qualitative
explanation in the frameworks of the proposed model and single out the
main clues to its understanding.

\section{Model}

The model to be developed uses the basis of the molecular-kinetic
theory of solid solutions \cite{18,20, o1, o2} and the state diagram  of the
Pd-Er alloys (in the $cT$-plane, where $c$ is the atomic concentration of Er atoms, $T$ is
temperature) \cite{19}. Namely, we assume that difference in the
atomic radii and electro-chemical characteristics of Er and Pd results 
in instability of the homogeneous state of the given solid solution Pd--Er. 
The instability gives rise to domains $\mathcal{Q}_{\mathrm{Er}}$ with
an increased value  $c_{+}(T)$ of the Er concentration whose composition 
approximately corresponds to the intermetallic ErPd$_{7}$, the first
equilibrium phase in the state diagram after the primary solid
solution of Er atoms in the Pd-matrix \cite{19}. These domains are 
surrounded by the maternal phase $\mathcal{Q}_{\mathrm{Pd}}$ (primary
solid solution) with a lower value $c_{-}(T)$ of the Er concentration 
equal to the limit of Er solubility in the
Pd-matrix. In equilibrium for the regular solid solution the
relationship between the values $c_{+}(T)$ and $c_{-}(T)$ is specified 
by the ``arms rule'' \cite{18}. However crystalline defects 
 such as vacancies, interstitial atoms, their complexes,
dislocations, grain boundaries, and so on can trap Er atoms. 
Furthermore we cannot exclude that other phases more enriched with Er
atoms, ErPd$_{3}$, Er$_{10}$Pd$_{21}$, etc., appear in the neighborhood of
these defects. A multiscale
hierarchical structure can form itself in this system in general.

\begin{figure}
\begin{center}
\includegraphics[scale=1]{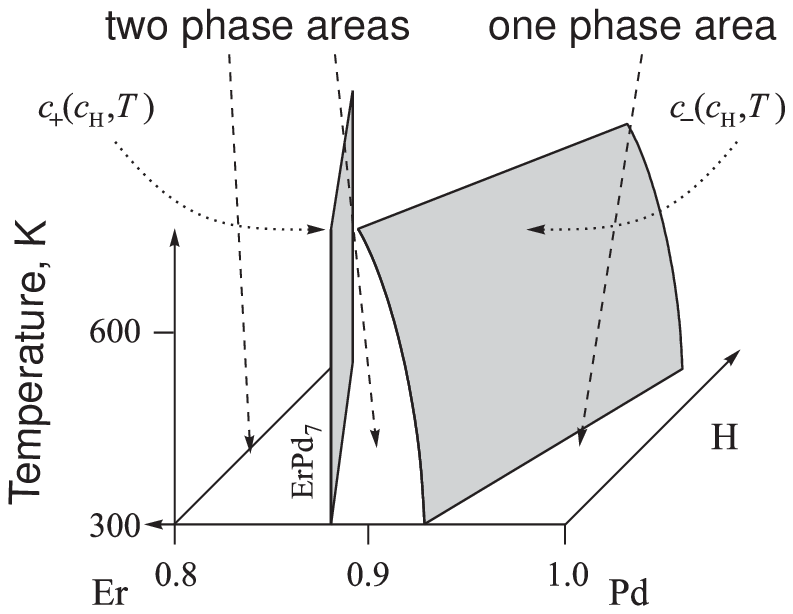}
\end{center}
\caption{Supposed state diagram for the triple system Pd--Er--H in the
region of small Er atom concentration. In drawing this diagram we have 
based on the results of the paper \protect\cite{19} for Pd--Er system
and on the  general
assumptions about the hydrogen effect on the state of the Pd--Er alloy.
\label{F1}}
\end{figure}

Penetrated into the alloy hydrogen atoms disturb the equilibrium in
two ways. First, the hydrogen saturation converting the solid
solution Pd--Er into the ternary system Pd--Er--H changes the phase
equilibrium in the metal subsystem. The higher affinity between 
H and Er in comparison with the Pd and H should give rise to the 
complexes \{Er--H\} that contain not only binary bonds Er--H but also
many bond structures, for example, Er--H--Er. The first type clusters,
Er--H, seem to be typical for regions with low values of the
Er concentration, whereas in the Er enriched phase
$\mathcal{Q}_{\mathrm{Er}}$  as well as near the 
solubility limit of Er atoms in the Pd-matrix the amount of the second 
type clusters should be essential. The presence of the Er--H--Er bonds 
effectively makes the interaction of Er atoms with each other
stronger, which has to cause the state diagram of the triple solution 
Pd--Er--H take a form shown at fig.~\ref{F1}. Drawing this diagram
we were keeping in mind one presented in \cite{19} for the binary
solution Pd--Er and treating hydrogen atoms as a source of the
additional Er--Er attraction. Let us explain the proposed form of this 
diagram in more details. Under the adopted assumptions the instability 
of the system homogeneous state is to occur at lower concentrations of
Er atom than it would be in the absence of hydrogen
atoms. Correspondingly, the solubility limit $c_{-}(c_{\mathrm{H}},T)$
of Er atoms in the Pd-matrix has to depend on the hydrogen
concentration  decreasing with the value of
$c_{\mathrm{H}}$ grows. In its turn, the composition of the 
intermetallic ErPd$_{x(\mathrm{H})}$ or, what is the same, the value 
$c_{+}(c_{\mathrm{H}},T)$ seems also to depend on the local values of
the hydrogen concentration $c_{\mathrm{H}}$ going into the region of
higher value with $c_{\mathrm{H}}$ grows. We assume the intermetallic
ErPd$_{x(\mathrm{H})}$ to originate from ErPd$_{7}$ and to form a  
certain phase in the nonequilibrium system under consideration. We
note that the concept of true intermetallic concerns a rigorously
ordered phase with specific stoichiometric composition. So the concept
of intermetallic for nonequilibrium systems requires an individual
consideration. For given system the experimental data demonstrate the
presence of a certain phase containing hydrogen atoms where the 
concentration of Er atoms closely matches the composition of the
intermetallic ErPd$_{7}$. However, whether this phase can be treated 
as a true intermetallic ErPd$_{7}$ or it comprises its small clusters,
equilibrium or nonequilibrium, containing hydrogen atoms is of low
importance in the present study. We need only the presense of a certain
quasiequilibrium phase experimentally observed, where the atomic 
concentration of Er atoms is about 10\%. We suggest this phase to be 
relative to the true intermetallic ErPd$_{7}$ so below we also will
refer to it as to  ``intermetallic'' enclosing with quotation marks.

Second, due to higher affinity between hydrogen and the defects in
comparison with the regular structure of the Pd-matrix hydrogen atoms
have to be attracted by defects containing also an increased amount of 
Er atoms. By the same reasons as discussed in the previous paragraph
hydrogen atoms penetrating into the defects located in the phase 
$\mathcal{Q}_{\mathrm{Pd}}$ have to density the Er distribution in them,
and, so, these defects gain the ability to absorb new Er atoms. 
In another words, hydrogen atoms penetrating  into phase  
$\mathcal{Q}_{\mathrm{Pd}}$ increase the  ``capacity'' of defects
being activated by them as traps or sinks of Er atoms. Hydrogen atoms
penetrating into
the  ``intermetallic'' $\mathcal{Q}_{\mathrm{Er}}$ also affect the state 
of its defects make variations in the ``intermetallic''
composition possible. In fact, without hydrogen  the composition
of ErPd$_7$ is fixed (in the state diagram the intermetallic  ErPd$_7$ 
corresponds to a line rather than two-dimensional region
\cite{19}). So, defects located in this intermetallic are to contain
other phases enriched with Er atoms to higher degrees, e.g., 
ErPd$_3$. Now, these local phases gain the possibility to dissolve in
the ``intermetallic'' $\mathcal{Q}_{\mathrm{Er}}$ injecting new Er atoms 
into in it. In other words, the hydrogen saturation into the phase
$\mathcal{Q}_{\mathrm{Er}}$ activates its defects as local sources of Er 
atoms.

The interface $\mathcal{T}$ separating the phases
$\mathcal{Q}_{\mathrm{Pd}}$ and $\mathcal{Q}_{\mathrm{Er}}$ is of atomic
size. Therefore during the relaxation course this interface locally
must be in quasiequilibrium, i.e. the concentration of Er atoms  in the 
regions of the phases $\mathcal{Q}_{\mathrm{Er}}$ and
$\mathcal{Q}_{\mathrm{Pd}}$ adjacent to the interface $\mathcal{T}$ must 
be equal to $c_{+}(c _{\mathrm{H}},T)$ and $c_{-}(c_{\mathrm{H}},T)$,
respectively. This quasiequilibrium is the result of practically
instantaneous redistribution of Er atoms on atomic scales in the
vicinity of the interface $\mathcal{T}$. The following diffusion of Er 
atoms into the bulk of the phases $\mathcal{Q}_{\mathrm{Er}}$ and
$\mathcal{Q}_{\mathrm{Pd}}$ tends to make their individual composition
uniform, causing the interface $\mathcal{T}$ to move. In fact,
generally the diffusion flux at the interface $\mathcal{T}$ on the
side of the phase $\mathcal{Q}_{\mathrm{Pd}}$ and that of the phase
$\mathcal{Q}_{\mathrm{Er}}$ is different. Thus, due to atom conservation 
the interface $\mathcal{T}$ is to move to compensate this difference  
giving rise to an effective source of Er atoms proportional to  
$v[c_{+}(c_{\mathrm{H}},T)- c_{-}(c_{\mathrm{H}},T)]$, where $v$ is
the normal velocity of the interface $\mathcal{T}$.

\begin{figure}
\begin{center}
\includegraphics[scale=1]{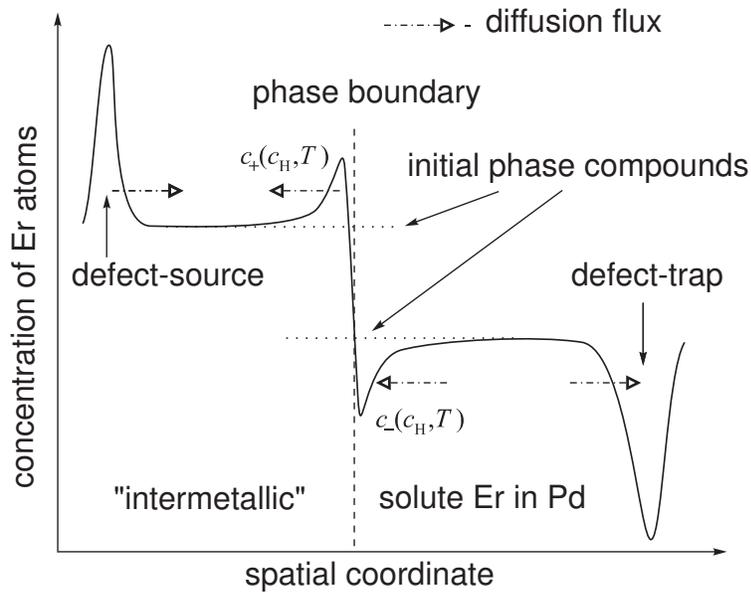}
\end{center}
\caption{Illustration of Er distribution at the beginning of the
relaxation process.\label{F2}}
\end{figure}

Nonmonotonic motion of the interface $\mathcal{T}$ being the essence
of the observed relaxation process is due to change in the profile of
Er distribution as time goes on. Fig.~\ref{F2} sketches the Er
distribution at the initial stage of relaxation after the hydrogen saturation. 
Light arrows point the diffusion direction of Er atoms in phases 
$\mathcal{Q}_{\mathrm{Pd}}$ and $\mathcal{Q}_{\mathrm{Er}}$. Whence it
follows that for each of the phases individually the change in the 
Er equilibrium concentration near the
interface $\mathcal{T}$ and the presence of the activated defects
cause the diffusion flux in the opposite directions.   
At the initial stage the distant defects cannot affect the interface
motion, so, it is governed by the change in the equilibrium state only.
As time goes on the diffusion flux induced by the activated defects
can become dominant and the interface $\mathcal{T}$ will move in the
opposite direction.

We suppose that this phenomenon is a backbone of the nonmonotonic
relaxation induce by the hydrogen saturation into systems like the Er--Pd
alloy. 

Before passing to a specific mathematical model we would like to note
the following. As seen from the present analysis the individual role
of the two phases is approximately the same. The latter enables us to
confine our consideration to only one of them, for example, the phase   
$\mathcal{Q}_{\mathrm{Pd}}$. Taking into account both of them 
only numerical coefficients in the expressions to be obtained can be changed, 
makes no sense at the current stage of the theory development.

\section{Master equations}

Taking into account aforesaid we write the diffusion equation for 
the atomic concentration $c$ of Er atoms in the phase 
$\mathcal{Q}_{\mathrm{Pd}}$ regarded as the half-space $z>0$ (we have 
attached the coordinate system to the interface $\mathcal{T}$ moving at 
the velocity $v$ in the laboratory frame): 
\begin{equation}
\frac{\partial c}{\partial t}-v\frac{\partial c}{\partial z}=D\frac{\partial
^{2}c}{\partial z^{2}}-\frac{1}{\tau_{\mathrm{tr}}}qcl\delta (z-z_{\mathrm{tr}})\,,
\label{1}
\end{equation}
where $D$ is the diffusion coefficient of Er atoms treated as a
constant and the second term in the right-hand side of equation~(\ref{1})
describes the Er atom trapping by a defect approximated by the
$\delta$-function located at a point $z_{\mathrm{tr}}$. This trap-defect is
characterized by the following parameters, $l$ is its real physical size,
$q$ is the capacity, i.e. the number of free seats for Er atoms at the
current time, and $\tau_{\mathrm{tr}}$ is the characteristic time
during which, on the average, the defect traps an Er atom located in
its neighborhood (not to be confused with the lifetime of Er atoms
inside the defects regarded infinitely long in the given model).  
In the frame $\{z,t\}$ attached to the interface $\mathcal{T}$
the trap-defect  moves towards the interface at the velocity $v$, i.e. 
\begin{equation}
\frac{dz_{\mathrm{tr}}}{dt}=-v\,.
\label{2}
\end{equation}
Here $z_{\mathrm{tr}}^{0}$ is the initial position of the defect (at $t=0$)
and is one of the model parameters.   
Since the newly trapped atoms occupy vacant places inside the defect its
capacity decreases in time according to the equation:
\begin{equation} \frac{dq}{dt}=-\frac{qlc(z_{\mathrm{tr}},t)}
{a\tau_{\mathrm{tr}}}\,,
\label{3}
\end{equation}
and $q_{0}$, the initial capacity of the trap-defect activated by the
hydrogen saturation, is also a model parameter ($a$ is a period of the lattice).

At the interface $\mathcal{T}$ ($z=0$) the Er concentration is set
equal to: 
\begin{equation}
c_{i}=c_{0}-\Delta _{\mathrm{Pd}}\,,
\label{4}
\end{equation}
where $c_{0}$ is the initial concentration of Er atoms in the phase
$\mathcal{Q}_{\mathrm{Pd}}$ before the hydrogen saturation and 
$\Delta_{\mathrm{Pd}}$ is a change in this concentration caused by the
hydrogen effect on thermodynamical equilibrium of the phase
boundary. Ignoring diffusion of Er atoms inside the phase 
$\mathcal{Q}_{\mathrm{Er}}$ we can represent the law of Er atom
conservation at the interface $\mathcal{T}$ in the form 
\begin{equation} \Delta v=D\left.
\frac{\partial c}{\partial z}\right| _{z=0}\,, 
\label{4a}
\end{equation}
where $\Delta =c_{+}(c_{H},T)-c_{-}(c_{H},T)$ (fig.~\ref{F2}). 

At the initial fast state of the system relaxation only the phase
boundaries attain new equilibrium and the defect activation is
completed, whereas the state of the phase bulk remains unchanged. The
latter enables us to adopt the following initial conditions for the Er
concentration \textit{inside} the  phase $\mathcal{Q}_{\mathrm{Pd}}$:
\begin{equation}
c(z,t)=c_{0}\quad \mathrm{for}\quad z>0,\ t=0\,.
\label{5}
\end{equation}

The system of equations (\ref{1})--(\ref{5}) forms the required  
relaxation model for Pd--Er--H system. Let us discuss its features in
detail. 

The proposed model contains four principal parameters:
\begin{equation}
\frac{\Delta }{c_{0}}\,,\quad \frac{\Delta _{\mathrm{Pd}}}{c_{0}}\,,\quad
\Omega =\frac{q_{0}lz_{\mathrm{tr}}^{0}}{D\tau _{\mathrm{tr}}}\,,\quad \Lambda
=\frac{c_{0}z_{\mathrm{tr}}^{0}}{q_{0}a}\,.
\label{6}
\end{equation}
For the sake of simplicity we rewrite system (\ref{1})--(\ref{5}) in the
dimensionless form by introducing the normalized variable $\eta
=c/c_{0}$ and $\theta=q/q_{0}$, and the dimensionless spatial and
temporal coordinates $\xi =z/z_{\mathrm{tr}}^{0}$
and $\tau =(Dt)/(z_{\mathrm{tr}}^{0})^{2}$:
\begin{equation}
\frac{\partial \eta }{\partial \tau }-\vartheta \frac{\partial \eta
}{\partial \xi }=\frac{\partial ^{2}\eta }{\partial \xi ^{2}}-\Omega
\theta \eta \delta (\xi -\xi _{\mathrm{tr}})\,,
\label{7}
\end{equation}
where $\xi _{\mathrm{tr}}=z_{\mathrm{tr}}/z_{\mathrm{tr}}^{0}$, and
$\vartheta =vz_{\mathrm{tr}}^{0}/D$ is the dimensionless velocity
of the interface $\mathcal{T}$. 
The scales of the spatial coordinate and time are chosen so that the unit
value of $\xi$ correspond to the defect position
($z_{\mathrm{tr}}^{0}$) at initial time and the unit value of $\tau$
characterize the diffusive penetration of Er atoms over scales about  
$z_{\mathrm{tr}}^{0}$.
Equation (\ref{7}) is subject to the boundary conditions:
\begin{equation}
\left. \eta \right| _{\xi =0}=1-\frac{\Delta _{\mathrm{Pd}}}{c_{0}}\,,\quad
\left. \frac{\partial \eta }{\partial \xi }\right| _{\xi =0}=\frac{\Delta
}{c_{0}}\vartheta
\label{8}
\end{equation}
and to the initial condition at $\tau =0$:
\begin{equation}
\eta (\xi )=1\quad \mathrm{for}\quad \xi >0\,.
\label{9}
\end{equation}
In the chosen frame the ``motion'' of the trap-defect 
$\xi_{\mathrm{tr}}(\tau)$ and its capacity $\theta (\tau )$ obey the equations:
\begin{equation}
\frac{d\xi _{\mathrm{tr}}}{d\tau }=-\vartheta \quad \mathrm{and}\quad
\frac{d\theta }{d\tau }=-\Lambda \Omega \theta \eta (z_{\mathrm{tr}},\tau)\,.
\label{10}
\end{equation}

If we ignore the influence of the defect on the interface motion  
and on the Er atom distribution near $\mathcal{T}$ (which is well
justified at the beginning of the relaxation process for  $\tau\ll 1$)
then equation~(\ref{7}) under conditions~(\ref{8})-(\ref{9}) admits
the automodel solution:
\begin{equation}
\eta ^{\ast }(\xi ,\tau )=1-\sqrt{\pi }\frac{\Delta }{c_{0}}\vartheta
_{0}\exp \left( \vartheta _{0}^{2}\right) \left[ 1-\func{erf}\left(
\frac{\xi }{2\sqrt{\tau }}+\vartheta _{0}\right) \right] \,,
\label{11}
\end{equation}
where parameter $\vartheta _{0}$ is the root of the transcendental equation
\begin{equation}
\vartheta _{0}\exp \left( \vartheta _{0}^{2}\right) \left[
1-\func{erf}\left( \vartheta _{0}\right) \right] =\frac{\Delta
_{\mathrm{Pd}}}{\sqrt{\pi }\Delta }
\label{12}
\end{equation}
and the interface velocity is given by the ratio $\vartheta=\vartheta
_{0}/\sqrt{\tau }$. Numerical analysis has shown, in particular, that for $\Delta
/c_{0}=1.5$ and $\Delta_{\mathrm{Pd}}/c_{0}=0.5$ the value $\vartheta_{0}\approx 0.2$
and on time scales $\tau \lesssim 1$ the interface velocity $\mathcal{T}$ 
is also about 1. Therefore, on one side, for the given values of the parameters 
the interface $\mathcal{T}$ is to reach the defect in time $\tau$ being about 10.

\begin{figure}
\begin{center}
\includegraphics[scale=1]{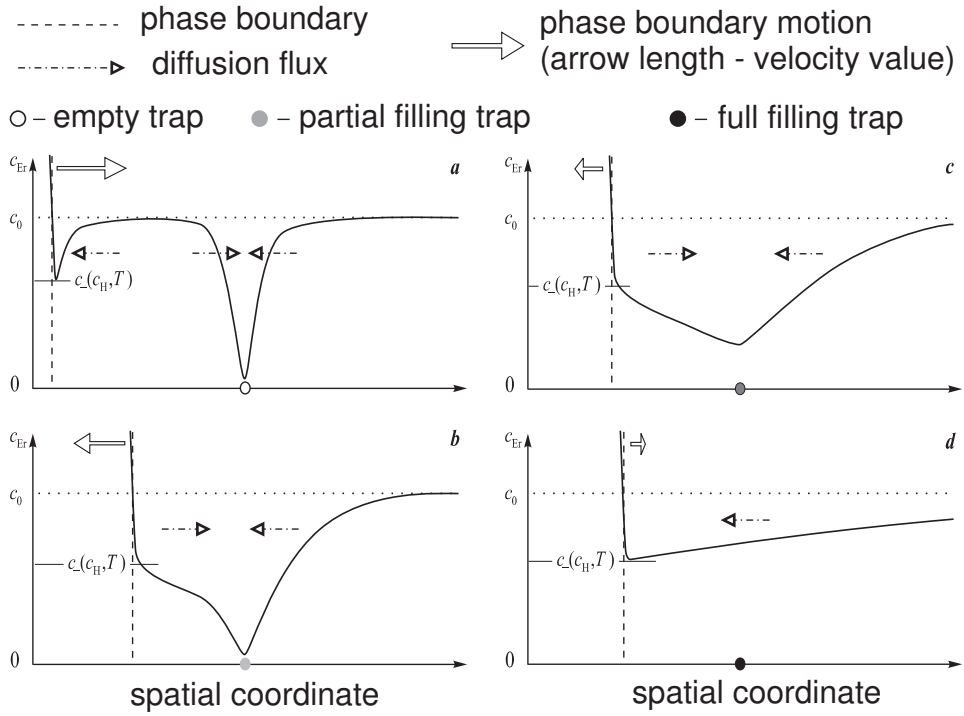}
\end{center}
\caption{Illustration of  evolution of Er distribution inside the phase 
$\mathcal{Q}_{\text{Pd}}$  during the interface motion ($t_a < t_b <
t_c <t_d$). 
\label{F3}}
\end{figure}

On the other side, if the parameter $\Omega \gg 1$, because
the characteristic time $\tau_{\mathrm{tr}}$ of trapping Er atoms by
the defect is small, and its initial capacity $q_{0}$ is such that $\Lambda
\lesssim 1$ then, by virtue of equation~(\ref{7}), the defect will
trap all the Er atoms located near it within a sufficiently short time
$\tau\sim\Omega^{-2}$.  As the result the distribution of Er atoms in the
vicinity of the defect will take the form shown in fig.~\ref{F3}$a$ and it is
characterized by small values of variable  $\eta (z_{\mathrm{tr}}, \tau)$.
As follows from the inequality $\Omega \gg 1$ and equations~(\ref{10}) 
this form does not change until the interface $\mathcal{T}$ comes
sufficiently close to the defect or until the defect is filled up with
Er atoms. As it must, the width of the defect neighborhood depleted of
Er atoms increases in time as $\sqrt{\tau }$. Therefore, for
time scales $\Omega ^{-2}\ll \tau \ll 1$ and the defect neighborhood
of thickness about several $\sqrt{\tau }$ equation~(\ref{7}) gives  
the distribution of Er atoms approximately of the form:
\begin{equation}
\eta (\xi ,\tau )=\func{erf}\left( \frac{\left| \xi -\xi
_{\mathrm{tr}}\right| }{2\sqrt{\tau }}\right)
\label{12a}
\end{equation}
and, in addition, the dimensionless rate of Er atom trapping is
estimated as 
\begin{equation}
\int\limits_{\mathcal{Q}_{\mathrm{tr}}}d\xi \,\Omega \theta \eta \delta (\xi
-\xi _{\mathrm{tr}})\approx \Omega \eta (z_{\mathrm{tr}},\tau )\approx 2\left.
\frac{\partial \eta }{\partial \xi }\right| _{\xi =\xi
_{\mathrm{tr}}+0}\approx \frac{2}{\sqrt{\pi \tau }}\,,
\label{12b}
\end{equation}
Whence, in particular, setting $\eta (z_{\mathrm{tr}},\tau )\ll 1$ we 
have got the estimate of the time during which
distribution~(\ref{12a}) develops.

After a lapse of time about $\tau \lesssim 1$ the regions near the
interface $\mathcal{T}$ and the defect where the Er concentration
exhibits substantial spatial variations begin to overlap and the
resulting form of the Er distribution is shown on fig.~\ref{F3}$b$. 
It is essential that in the given case the gradient of the Er
concentration at the interface $\mathcal{T}$ changes the sign, causing 
the interface to move in the opposite direction. As time passes further
and the defect is filled with Er atoms its effect on distribution of
Er atoms becomes weaker and the Er concentration in its immediate
neighborhood starts to grow (fig.~\ref{F3}$c$). Then as follows from 
the second equation of (\ref{10}) and estimate~(\ref{12b}) after a time 
$\tau \sim \Lambda ^{-2}$ the defect has practically no effect on
diffusion of Er atoms and their distribution takes again the
automodel form~(\ref{11}) (fig.~\ref{F3}$d$). In this case the
gradient of Er concentration at the interface $\mathcal{T}$ changes
the sign again and the interface motion returns to the initial
direction.

\begin{figure}
\begin{center}
\includegraphics[scale=1]{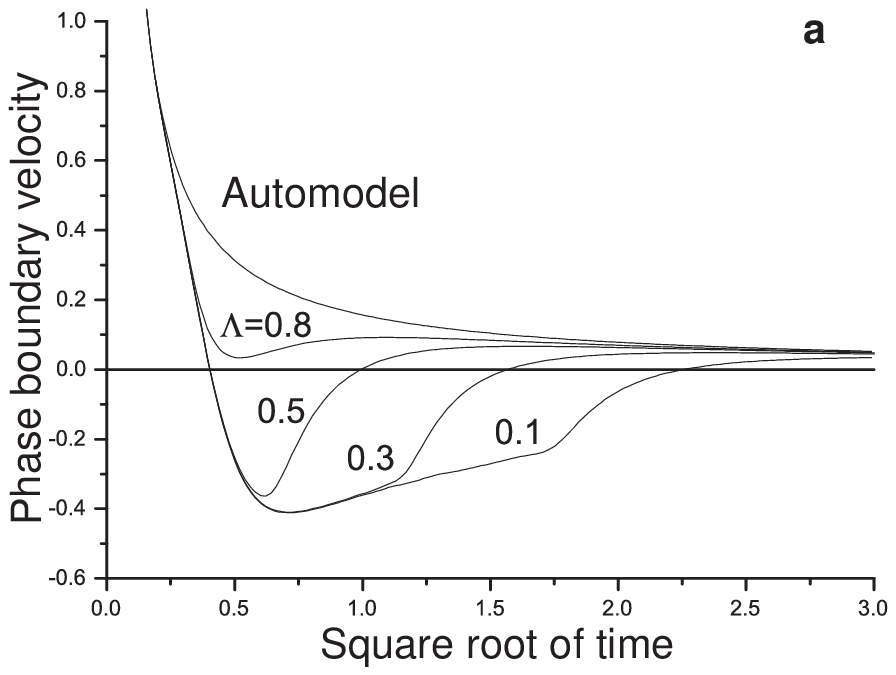}\par
\includegraphics[scale=1]{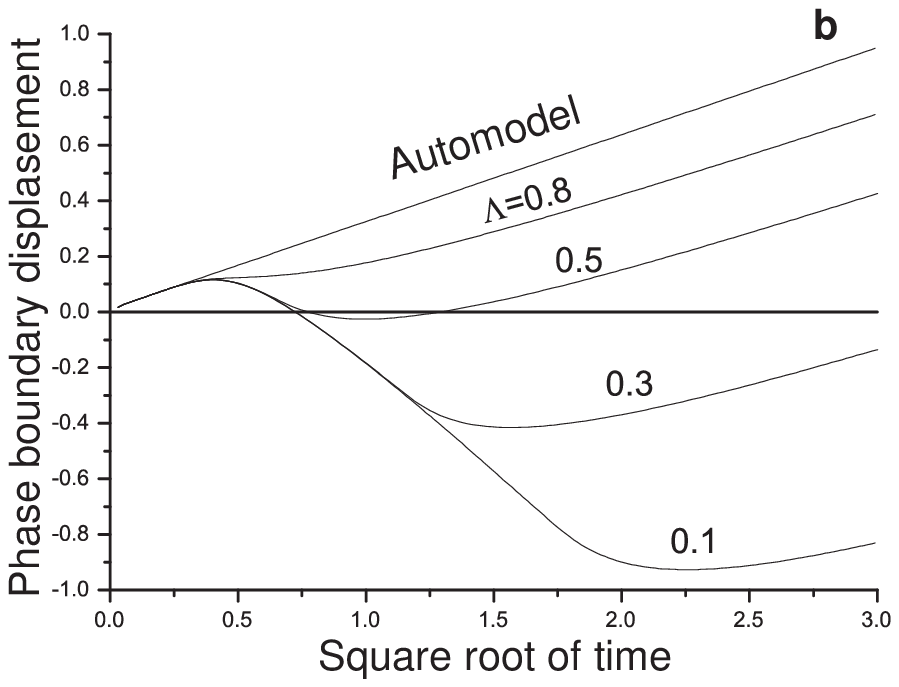}
\end{center}
\caption{Results of numerical simulation of the interface motion
(in dimensionless form):  
\textit{a)} the interface velocity $\vartheta$ vs. the square root of time
$\sqrt{\tau}$,
\textit{b)} the interface position, $\int_{0}^{\tau}
\vartheta(\tau')d\tau'$,
vs. square root of time $\sqrt{\tau}$.  
($\Delta /c_{0}=1.5$, $\Delta_{\text{Pd}}/c_{0}=0.5$, $\Omega=1000$)
\label{FN}}
\end{figure}

The described process is the basis of the proposed model for the
nonmonotonic relaxation in system Pd--Er--H. So, to substantiate our 
speculations we present also curves describing the interface motion
that have been obtained by solving equations~(\ref{7})--(\ref{10})
numerically. Fig.~\ref{FN} illustrates the interface motion for
various values of the defect capacity in the interval $0.1\leq\Lambda\leq
0.8$, with the other parameters being fixed, $\Delta /c_{0}=1.5$, 
$\Delta_{\mathrm{Pd}}/c_{0}=0.5$, and $\Omega=1000$.
When the defect capacity is large enough so the parameter $\Lambda< 0.8$
we see the nonmonotonic motion of the interface $\mathcal{T}$. 
We have taken the abscissa as the square root of time, $\sqrt{\tau}$,
to make the results more clear. It is due to the fact that the second stage 
of the interface motion, i.e., when the interface moves in the
direction opposite to the initial one, is characterized by the duration 
exceeding substantially the one of the first initial stage. It should be
noted similar behavior of the relaxation processes has been observed
experimentally \cite{8,12}, where each of the points of time
variations in the phase relative volumes changing sign is distant
from preceding one by a more and more prolonged time interval.
We point out once more that obtained results are of qualitative level 
and to compare quantitatively  the experimental data and theoretical
results a more sophisticated model is required.

\section{Acknowledgments}

The authors are grateful to  A.\,L. Udovsky for the discussion and useful remarks,
Russian Foundation of Basic Research for support, Grant 99-02-16135, 
and Universities of Russia, fundumental researching program, Grant 990156.

\newpage

\begin{thebibliography}{99}

\bibitem{1}\textit{Hydrogen in metals} (eds. G. Alefeld and J. Volki),
Moscow, Mir publ., 1981, 467~p.

\bibitem{2}V.\,M. Avdjukhina, A.\,A. Katsnelson, G.\,P. Revkevich, Surf. Invest.
(RSNI),\textbf{14}, n.~2, 30, 1999.

\bibitem{3}V.\,A. Goltzov, V.\,A. Lobanov, RAS USSR, \textbf{283}, n.~3, 598,
1985

\bibitem{4}G.\,P. Revkevich, A.\,A. Katsnelson, V.\,M. Khristov, Metallophisika,
\textbf{11}, n.~3, 57, 1989

\bibitem{5}A.\,A. Katsnelson, M.\,A. Knyazeva, A.\,I. Olemskoi, Phys.Sol.St., 
\textbf{41},n.~9, 1621, 1999

\bibitem{6a}E. Feder, \textit{Fractals}, Moscow, Mir publ., 1991, 260~p.

\bibitem{6b}M.\,B. Isichenko, Rev. Mod. Phys., \textbf{64}, n.~4, p.~961, 1992.

\bibitem{6c}V.\,S. Ivanova, A.\,S. Balankin, I.\,Zh. Bunin, A.\,A. Oksogoev,
\textit{Synergy and fractals in science of material}, Moscow, Nauka, 1994, 384~c.

\bibitem{7}A.\,I. Olemskoi, A.\,V. Khomenko, JETP, \textbf{110}, n.~6, 2144, 1996

\bibitem{8}V.\,M. Avdjukhina, A.\,A. Katsnelson, D.\,A. Olemskoi, A.\,I. Olemskoi,
G.\,P. Revkevich, Phys.Met.Met., \textbf{88}, n.~6, 63, 1999

\bibitem{9}A.\,A. Katsnelson, A.\,I. Olemskoi, I.\,V. Sukhorukova, G.\,P.
Revkevich, Physics-Uspekhi, \textbf{165}, n.~3, 331, 1995

\bibitem{10}V.\,M. Avdjukhina, A.\,A. Katsnelson, G.\,P. Revkevich, Crystallogr.Rep.,
\textbf{44}, n.~1, 49, 1999

\bibitem{12}V.\,M. Avdjukhina, A.\,A. Katsnelson, G.\,P. Revkevich, Khan Kha Sok and other, 
Alternative Energy and Ecology, \textbf{1}, n.~1, 11, 2000 

\bibitem{13}V.\,M. Avdjukhina, A.\,A. Katsnelson, N.\,A. Prokofiev, G.\,P.
Revkevich, Mosc.Univ.Phys.Bull., \textbf{3}, n.~2, 70, 1998

\bibitem{14}M. Myers, M.\,S. Baskes, H.\,K. Birnbaum et.al., Rev. Mod. Phys.
\textbf{64}, n.~2, 559, 1992

\bibitem{19}Z. Du, H. Yang, J. Alloys Comp. \textbf{299}, 199, 2000

\bibitem{15}E. Lorenz, J. Atmosph. Scienc. \textbf{20}, 130, 1963

\bibitem{16}A.\,I. Olemskoi, I.\,A. Sklyar, Physics-Uspekhi \textbf{162}, n.~6, 29, 1992

\bibitem{17}A. Damask, J. Dins, \textit{Point defects in metals}, Moscow,
Mir publ., 1966, 291~p.

\bibitem{18}A.\,A. Smirnov, \textit{Molecular-kinetic theory of metals}, 
Moscow, Nauka, 1966, 488~p. 

\bibitem{20}A.\,A. Katsnelson, A.\,I. Olemskoi, \textit{Microscopic 
Theory of Nonhomogeneous Structures}), 
Mir publ. (Moscow), AIP (New York), 1990, 357~p.

\bibitem{o1}K. Kawasaki, T. Ohta, Prog. Theor. Phys., \textbf{67}, 147, 1982

\bibitem{o2}K. Kawasaki, T. Ohta, Prog. Theor. Phys., \textbf{68}, 129, 1982

\end{thebibliography}
\end{document}